\documentstyle[sprocl,epsfig]{article}

\def\be{\begin{equation}}
\def\ee{\end{equation}}
\def\bea{\begin{eqnarray}}
\def\eea{\end{eqnarray}}

\begin{document}

\title{CALICE ELECTROMAGNETIC CALORIMETER READOUT STATUS\\}

\author{P.~D.~DAUNCEY, representing the CALICE collaboration}

\address{Imperial College London, Prince Consort Road, London SW7 2AZ, UK}


\maketitle\abstracts{
The status of the prototype readout boards for the
CALICE electromagnetic calorimeter is presented. Results on
linearity, noise, and minimum ionising signals, both from a radiactive
source and cosmic rays, are shown.
}
  
\section{Introduction}
The CALICE Electromagnetic CALorimeter (ECAL) will be a silicon-tungsten
sampling calorimeter~[1] 
with 30 layers of silicon wafers, each
containing an $18 \times 18$ array of diode pads, giving 9720 channels to
be read out. The signal from each pad is amplified by a Very Front End (VFE)
ASIC~[2] 
which multiplexes the signals from 18 pads onto one
output line. The silicon wafers and VFE chips are mounted on 
VFE PCBs~[3], 
each containing up to 12 VFE ASICs, equivalent to 216 pads. The signal
cables from the VFE PCBs connect directly to the readout boards and the
latter are the subject of this paper.

\section{The Readout Board Design}
The CALICE ECAL Readout Cards (CERC)~[4] 
are 9U,
double-sided, VME boards originally based on the CMS silicon tracker
Front End Driver board~[5]. 
They contain 16-bit ADCs for
each VFE ASIC output and 16-bit DACs for calibration. The 16-bit ADCs
can operate at up to 500\,kHz and so take around 80\,$\mu$s to read out
all 18 multiplexed channels from a VFE ASIC. There is no data reduction
on the CERCs; all signals from all channels will be read out and this
corresponds to 5\,kBytes per CERC per event, or 30\,kBytes total per event.
These data are stored in an on-board 8\,MByte memory, which allows over
a thousand events to be buffered during a spill before readout via VME.

\begin{figure}[ht]
     \begin{center}
     \begin{tabular}{cc}
     \mbox{\epsfig{file=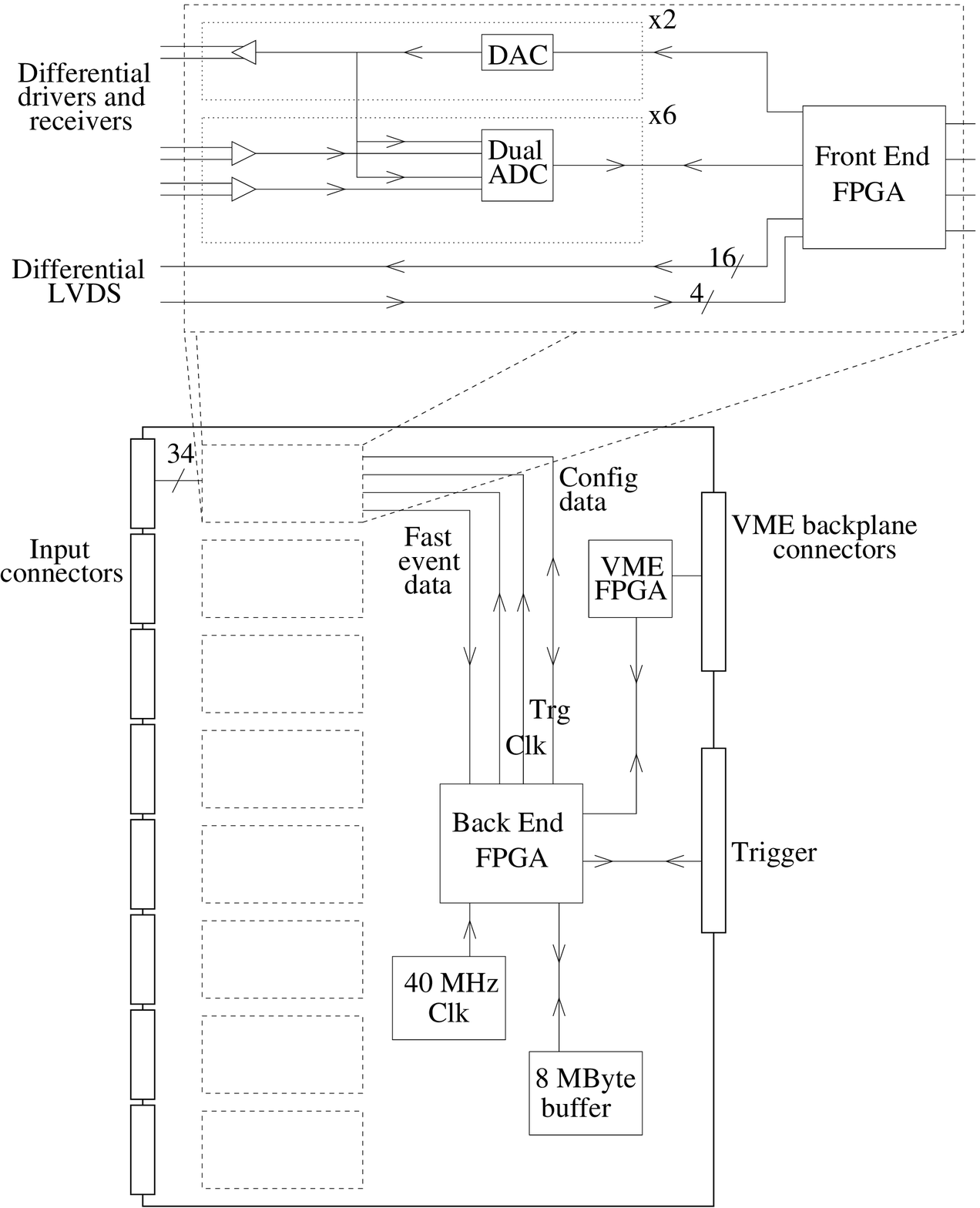,height=4cm}}&
     \mbox{\epsfig{file=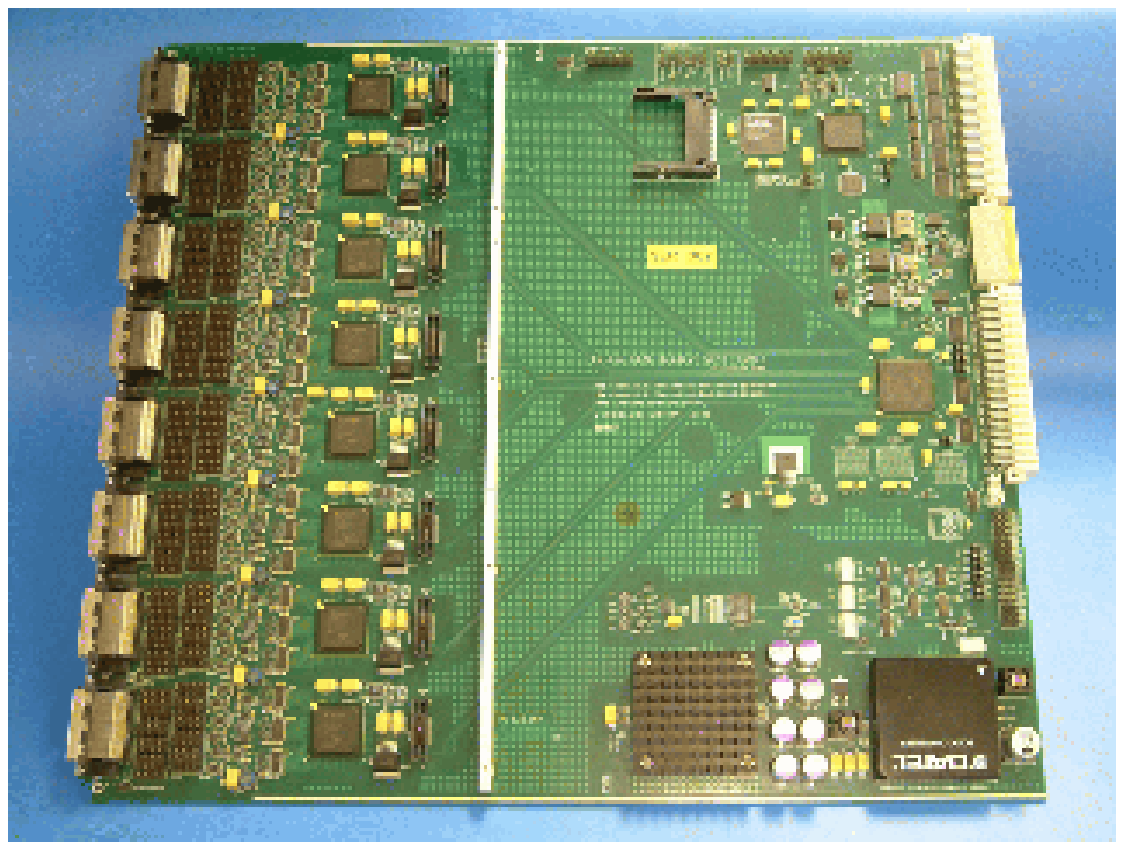,height=4cm}}
     \end{tabular}
     \end{center}
     \caption{Left: Overview of the CERC structure. Right: One of the two
prototype CERCs.}
     \label{fig:cerc}
     \end{figure}

An overview of the CERC is shown in figure~\ref{fig:cerc}.
Eight Front End (FE) FPGAs control all signals on the front panel connectors
to and from the VFE PCBs. The Back End (BE) FPGA gathers and buffers all
the event data from the FEs and provides the VME interface. Trigger logic
is implemented in the BE FPGA, for which the logic is only active on one
board. A total of six CERCs will be needed for the entire ECAL readout.
The prototype design of the CERCs was completed in summer 2003 and two
prototype CERCs were fabricated in November 2003. A photograph of one
of these prototypes is also shown in figure~\ref{fig:cerc}.

The CERC prototypes have been extensively tested with a prototype
VFE PCB containing one $6 \times 6$\,cm$^2$ silicon wafer.
Not all the final firmware for the FE and BE FPGAs is
yet complete; in particular, the complex data path from the FEs to
the 8\,MByte memory via the BE and then to the VME interface is not
functional. Hence, these tests have been done using a simple RS232
interface directly from the FEs to a PC. This gives a slow rate of
around 1\,Hz for readout (compared with an expected rate of around
1\,kHz for the final system) but this is sufficient for this round
of tests. In particular, the analogue parts (the ADC and DAC operation)
are operated in the same way
as for the final system, so results from these tests
give an accurate indication of the performance of these sections
of the CERC.


\section{Internal Loopback Tests}
In these tests, the DAC output was looped back directly to the ADC
inputs. This allowed a straightforward linearity scan of the system
independent of the VFE PCBs.
A simple
straight-line
fit gives residuals of around 1-2 ADC counts, corresponding to an
intrinsic linearity of 0.01\%. The gains, measured from the slope
of the ADC vs DAC response, show a spread of around 1\%. The noise
is around 1 ADC count for all channels. The
DAC was seen to saturate in the first 1\% of its range; this effect
is understood and can be corrected in the production version.

\section{Strontium Source Tests}
\begin{figure}[ht] 
     \begin{center}
     \vspace*{.2cm}
     \rotatebox{270}{%
\epsfig{file=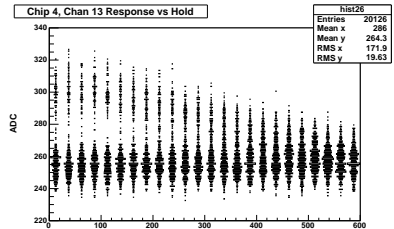,width=3cm}%
}
     \end{center}
     \caption{Pedestal and signal due to $^{90}$Sr as a function
of the sample-and-hold timing offset, in units of 6.25\,ns.} 
     \label{fig:sr}
     \end{figure}

A $^{90}$Sr beta source was used to give a relatively high rate of
signals in the wafers so as to determine the required timing. The
VFE sample-and-hold needs to be timed so as to capture the peak of
the CR-RC shape. This offset is software configurable in steps of
6.25ns. The signal due to the
$^{90}$Sr as a function of the timing offset is shown in figure~\ref{fig:sr}.
The separation of the signal (at around 300 counts) from the pedestal
(at around 255 counts) is clearly seen for small offsets, while the two
merge together as the offset increases. The offset was set to
maximise the separation of the signal from the pedestal.

\section{Cosmic Tests}
\begin{figure}[ht]
     \begin{center}
     \begin{tabular}{cc}
     \mbox{\rotatebox{270}{%
\epsfig{file=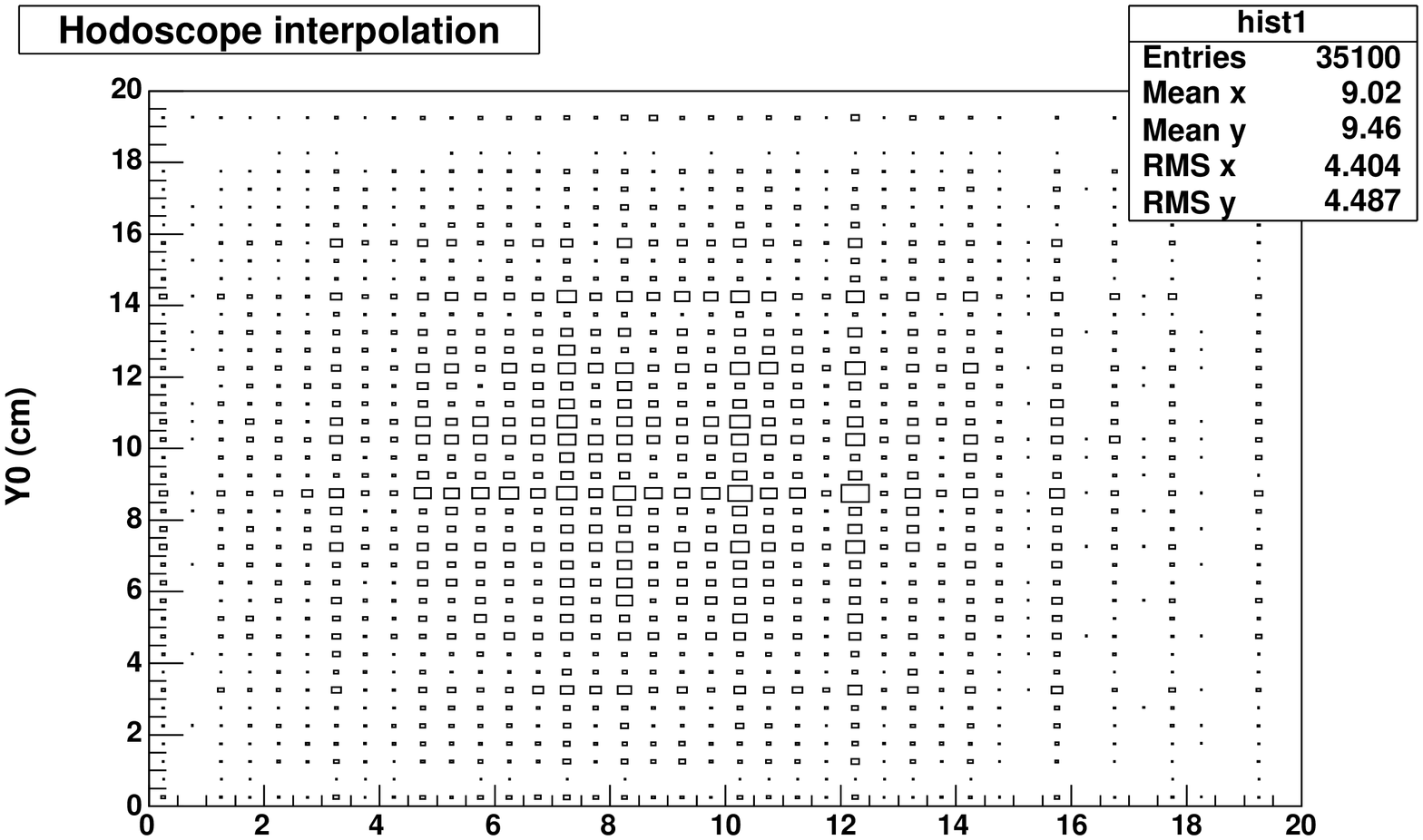,width=3cm}%
}}&
     \mbox{\rotatebox{270}{%
\epsfig{file=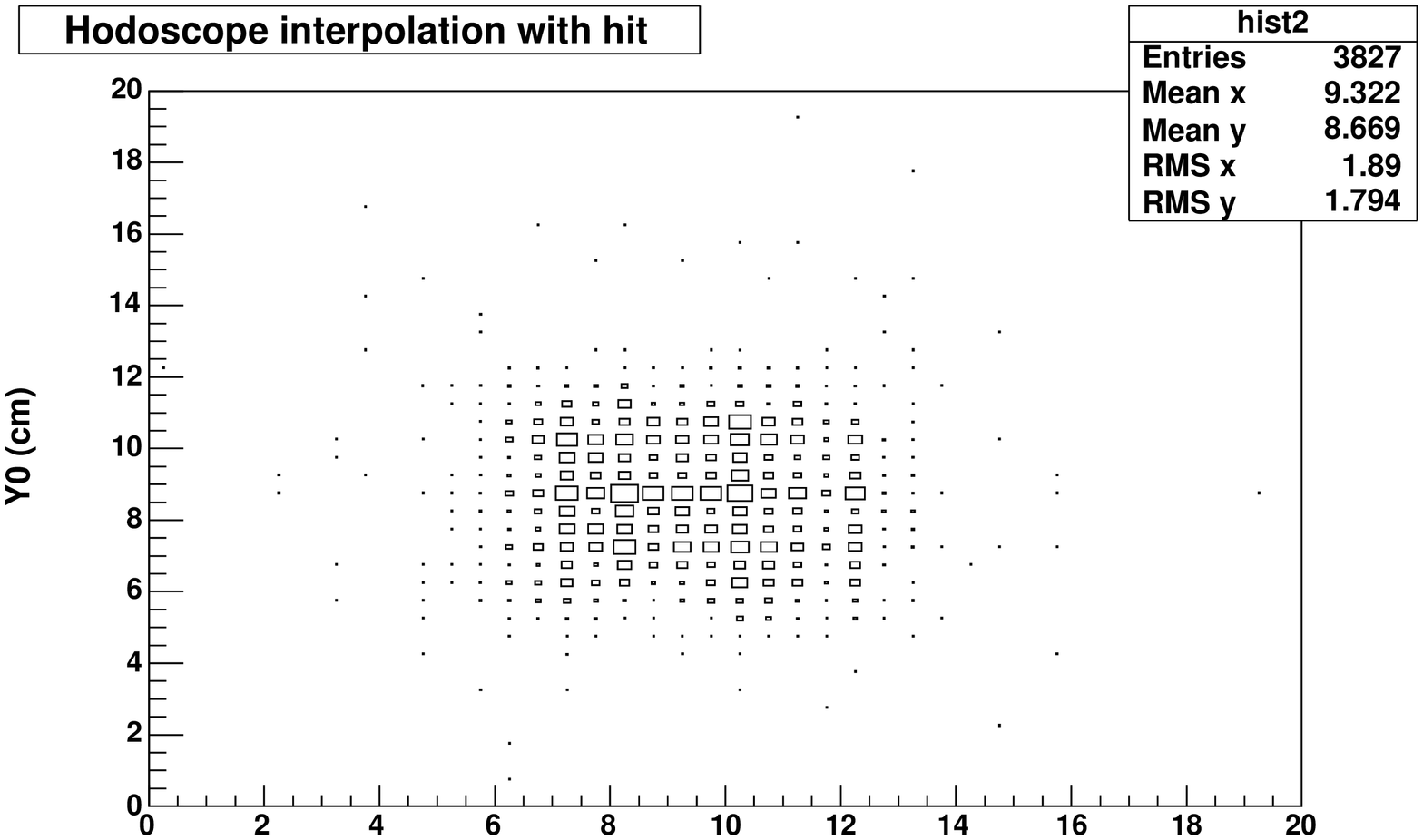,width=3cm}%
}}
     \end{tabular}
     \end{center}
     \caption{Interpolated position in cm
of hodoscope cosmic tracks in the
plane of the silicon wafer for (left) all events and (right) events
with any pad more than 40 ADC counts above pedestal.}
     \label{fig:hod}
     \end{figure}

A hodoscope scintillator cosmic ray telescope built at Ecole
Polytechnique~[3] 
was used to observe cosmic ray tracks in the
wafer. The tracks determined from
hodoscope information were interpolated into the plane of the
wafer and the distribution of these positions in the plane
is shown in figure~\ref{fig:hod}. The same distribution after requiring
any wafer pad to have a signal at least 40 ADC counts above pedestal
is also shown in figure~\ref{fig:hod}, clearly indicating
the outline of the
wafer in the hodoscope.

\begin{figure}[ht] 
     \begin{center}
     \vspace*{.2cm}
\rotatebox{270}{%
     \epsfig{file=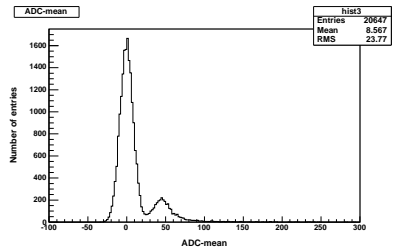,width=3cm}%
}
     \end{center}
     \caption{Distribution of pedestal-subtracted ADC values for
silicon wafer pads consistent with the hodoscope cosmic track
interpolation.}
     \label{fig:mip}
     \end{figure}

Pads consistent with the interpolated hodoscope track were used to
determine the cosmic signal size.
The distribution of ADC values relative to the pedestal for these
pads is shown in figure~\ref{fig:mip}.
The clear peak at around 45 ADC counts gives the average signal value
of a minimum ionising particle (MIP). This
means the full ADC
range corresponds to around 700 MIPs, which is sufficient for
the CALICE beam test studies. In addition, the MIP/noise
value is around 5/1, above the requirement of 4/1. Hence, the system
satisfies the basic requirements for CALICE.

\section{Future Plans}
Tests will continue over the summer with production of the
CERCs scheduled for early autumn. These
will be used for a first beam test of the whole ECAL
at DESY in a 6\,GeV electron beam
and this is scheduled for late
2004 and early 2005. Following this, further beam tests using
hadrons are foreseen, where the ECAL will be tested together
with several hadron calorimeters. These should continue throughout
2005 and into 2006.

\end{document}